# The direct test of the absence of the "quantum vampire's" shadow with use of thermal light

K. G. Katamadze,[1,2,3,4,*] E. V. Kovlakov,[1,2] G. V. Avosopiants,[1,2,5] and S. P. Kulik[1,2]

[1]*Quantum Technology Centre, M. V. Lomonosov Moscow State University,119991, Moscow, Russia*
[2]*Faculty of Physics, M. V. Lomonosov Moscow State University,119991, Moscow, Russia*
[3]*Institute of Physics and Technology, Russian Academy of Sciences,117218, Moscow, Russia*
[4]*National Research Nuclear University MEPhI,115409, Moscow, Moscow, Russia*
[5]*National Research University of Electronic Technology MIET, 124498, Moscow, Russia*

*Corresponding author: kgk@quantum.msu.ru*

**Counterintuitive nature of quantum physics leads to a number of paradoxes. One of them is a "quantum vampire" effect [1] consisting in the fact, that photon annihilation in a part of a large beam doesn't change the shape of the beam profile (i. e., doesn't cast a shadow), but may change the total beam intensity. Previously this effect was demonstrated just in a simplified double-mode regime [1,2]. In the current paper the direct test of shadow absence after the photon annihilation has been performed with use of thermal state of light at the input.**

Since quantum physics laws often contradict to our everyday experience, this science is really counterintuitive and full of paradoxes. Nevertheless, we try to use our intuition together with various analogies in order to understand some quantum phenomena. Sometimes it helps, but occasionally it leads to wrong statements.

In the current paper we consider the most fundamental quantum operations: photon creation and annihilation. Both of them have been probabilistically implemented in the experiments [3–5] and form a perfect quantum engineering toolbox [6,7]. They are useful not only for the probing of basic commutation relations [8–10], but also for noiseless amplification [11], preparation of strong Kerr nonlinearity [12], conditional entanglement [13] and so on. Such toolbox may be utilized both for fundamental studying of quantum thermodynamics [14,15] phenomena and for practical tasks of quantum metrology [16–18] and quantum computing [19,20]. Bellow we will focus on just the photon annihilation (or photon subtraction) process, but actually the most of the following consideration can be transferred to the photon creation process too.

The basic property of the photon annihilation operator is that it decreases the number of photons $n$: $\hat{a}|n\rangle = n|n-1\rangle$, so it is easy to represent the photon subtraction as some kind of nonlinear losses. This representation is intuitively obvious but completely wrong.

First, one can note, that photon annihilation not always leads to the decreasing of the mean photon number $\langle n \rangle$: it doesn't change the coherent state: $\hat{a}|\alpha\rangle = \alpha|\alpha\rangle$ and, moreover, it can increase $\langle n \rangle$. For example, the photon subtraction applied to thermal states of light leads to a doubling of $\langle n \rangle$ and subsequent photon annihilation is followed by the mean photon number linear growing [21–23]. Actually, it can be shown, that the second-order correlation function is calculated as $g^{(2)} = \langle n \rangle_s / \langle n \rangle_0$ [23], where $\langle n \rangle_0$ is an initial mean photon number and $\langle n \rangle_s$ is a mean photon number after the photon subtraction. This means, that photon annihilation decreases the mean photon number of non-classical states of light with $g^{(2)} < 1$ and increases the same quantity for classical ones with $g^{(2)} > 1$.

Second, in contrast to the regular optical losses, photon subtraction doesn't change the spatial intensity distribution. This is the basis of the "quantum vampire" effect [1], which is the subject of this Letter. Let us consider a small beam splitter placed into the large light beam (Fig. 1a). If the reflection coefficient is big enough, we will see a shadow (a dip) in the intensity distribution, measured by the camera. Now, let us choose the input beam power in such a way that the mean photon number in the reflected channel $\langle n_R \rangle = 1$ and then decrease the beam splitter reflectivity to a very small value: $r \ll 1$, $\langle n_R \rangle \ll 1$ (Fig. 1b). In this case no shadow can be observed, but next we may select the time events, when a photon in the reflection channel is detected, and use a detector click as a trigger for the camera. So one can measure an intensity distribution under condition that the photon number in the reflection channel $n_R = 1$ (Fig. 1c). This situation looks very similar to the previous one shown in Fig. 1a, and one may intuitively conclude, that the conditional intensity distribution will show the shadow presence again, but accurate consideration of this problem reveals, that the initial intensity distribution will be uniformly reduced for sub-Poissonian light or decreased for super-Poissonian light. This absence of the shadow has been called a "quantum vampire" effect.

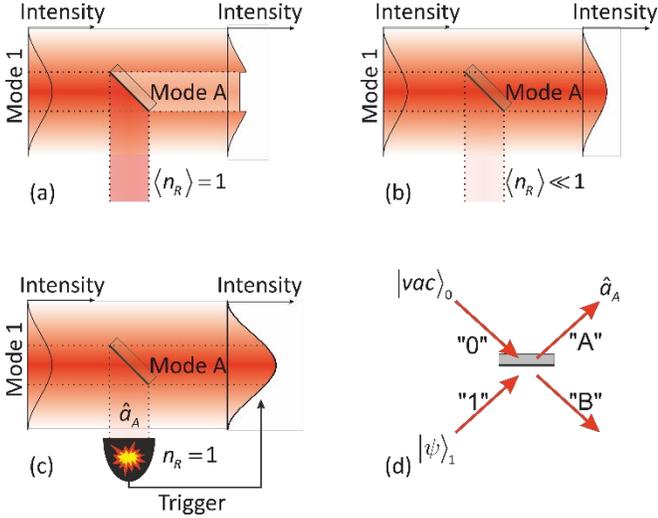

Fig. 1. The idea of the "quantum vampire" effect. (a) High-reflective beam splitter reflects 1 photon on the average $(\langle n_R \rangle = 1)$, introduces optical loss in a part of the beam (mode A) and casts a shadow in the output intensity distribution. (b) Low-reflective $(\langle n_R \rangle \ll 1)$ beam splitter doesn't cast any shadow. (c) Single-photon detector registers the reflected photons and allows one to measure the intensity distribution under condition $n_R = 1$, which corresponds to the photon subtraction in the mode A. It doesn't lead to the shadow, but changes the total intensity of the beam. (d) Simplified two-mode scheme of the quantum vampire effect.

To prove the last statement, we can note, that the combination of a low-reflective beam splitter and a single photon detector is a conditional implementation of the photon annihilation operator [4,5]. Let's consider a quantum state of light $|\psi\rangle_1$, directed to the input "1" of the beam splitter, the vacuum state $|vac\rangle_0$, directed to the input "0" and the photon subtraction in the output mode "A" (Fig. 1d). The annihilation operator in the mode A $\hat{a}_A = t\hat{a}_0 + r\hat{a}_1$, where $t$ and $r$ are the beam splitter transmission and reflectivity. Therefore, the action of the operator $\hat{a}_A$ on the state $|\psi\rangle_1$ can be written as follows:

$$\hat{a}_A |\psi\rangle_1 |vac\rangle_0 = \underbrace{t\hat{a}_0 |\psi\rangle_1 |vac\rangle_0}_{0} + r\hat{a}_1 |\psi\rangle_1 |vac\rangle_0,$$

i. e. that the photon subtraction in the mode "A" $\hat{a}_A |\psi\rangle_1$ leads to the subtraction in the initial mode "1" $r\hat{a}_1 |\psi\rangle_1$. Now we can denote the initial beam mode in Fig. 1 as a mode "1" and a part of the beam incident on the beam splitter as a mode "A" and thus we could conclude that the photon subtraction in a part of the beam leads to the photon subtraction in the whole beam.

Initially quantum vampire effect was demonstrated with Fock states of light distributed between two polarization modes [1], next it was tested with a thermal state distributed between two separated spatial modes [2], and in the current work we perform the initial shadow-measurement experiment (Fig. 1c) with a thermal light at the input. The scheme of the experimental setup is shown in Fig. 2.

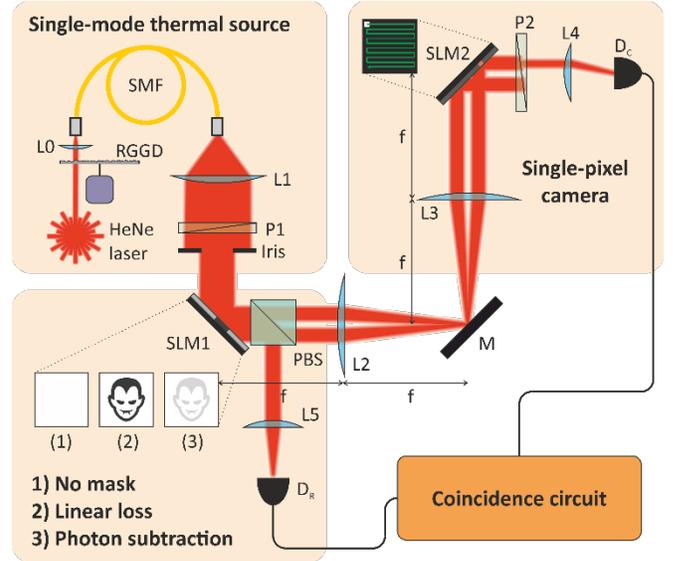

Fig. 2. The sketch of the experimental setup. L0 – L5 are lenses, RGGD – rotating ground glass disk, SMF – single-mode optical fiber, P1, P2 – polarizers, PBS – polarization beam splitter (P1, P2 and PBS transmit horizontal polarization), SLM1,2 – spatial light modulators (black pixels rotate polarization at 90 degree), M – mirror, $D_R$, $D_C$ – single-photon detectors.

The key element of the setup is a spatial lite modulator (SLM) produced by Cambridge Correlators. In general, the SLM is a liquid crystal matrix, which applies a user-defined pixel-dependent phase shift on the incident wave front. Since SLM is a polarization-depended device, it changes only the phase of diagonally polarized light. Thus the SLM sandwiched between two horizontally or vertically oriented polarizers works as a perfect amplitude modulator [24] (actually, that is the way, how ordinary liquid crystal display operates).

The setup has three parts: 1. Preparation of the initial single-mode thermal beam; 2. Beam transformation (applying optical losses or photon subtraction); 3. Output profile measurement performed by a single pixel camera [25] operated in a raster scanning mode.

The thermal light was prepared by passing cw HeNe laser radiation throw a rotating ground glass disk (RGGD) [26,27]. For the single mode selection it was coupled by the lens L0 to the single-mode optical fiber SMF. The fiber output beam was collimated by the lens L1, passed throw a horizontally oriented polarizer P1 and then its central near-uniform part was cut out by the iris.

The SLM1 was used to rotate the polarization of the input beam (white pixels correspond to horizontal polarization output and black one – to the vertical). In combination with a polarization beam splitter PBS, which transmits horizontal polarization and reflects vertical one, SLM1 worked as a beam splitter with a various reflection coefficient. We used three types of masks displayed on the

SLM1: (a) white mask which corresponded to the absence of the beam splitter, (b) high-contrast mask for linear loss (Fig. 1a) and (c) low-contrast mask acting as the low-reflective beam splitter (Fig. 1b, c). Single-photon detector $D_R$ placed in the reflection channel was used for the measurement of the mean photon number $\langle n_R \rangle$ in the case (b) and triggered the single-pixel camera in order to realize single-photon subtraction in the case (c).

The lens system L2, L3 combined with the mirror M formed the image of the SLM1 on the SLM2, which was the main part of the single-pixel camera. A white area of 11x11-pixels width (i. e. superpixel) on the black background applied to the SLM2 picked out the part of the image transmitted throw the polarizer P2 and collected by the single-photon detector $D_C$. Thus, we were able to perform the raster scanning of the image by changing the superpixel position. In the photon subtraction mode, the coincidence circuit selected camera signal ($D_C$ pulses) corresponding to the photon subtraction times ($D_R$ pulses).

Since our setup worked in cw regime, the time mode of all the measured states was defined by the detection scheme and its width was equal to 12 ns, which was much smaller than the thermal state correlation time. The total acquisition time of 3 seconds was used for each superpixel measurement.

The experimental results are presented in Fig. 4. Initial beam profile (Fig. 4a) was measured while the white mask (1) was applied on the SLM1. It looks as an almost a uniform ellipse with a bright contour caused by the beam diffraction on the iris. In Fig. 4b we present the profile of the beam transmitted throw the high-contrast vampire mask (2). The clear vampire image (shadow) can be seen. The mean photon number in the reflection channel $\langle n_R \rangle$ has been measured with $D_R$ and equals 1 (detector quantum efficiency and transmission losses have been accounted). In Fig. 4c we present the unconditional profile of the beam transmitted throw the low-contrast vampire mask (3). In this case $\langle n_R \rangle = 0.13$ and no shadow can be distinguished. Finally, in Fig. 4d we present a conditional beam profile corresponding to the photon subtraction times (when $n_R = 1$) formed by coincidence counts. Despite the fact that this situation is intuitively similar to the previous case presented in Fig. 4b, it is impossible to distinguish any shadow in Fig. 4d. On the other hand, one can note (see color bar), that the whole profile becomes twice brighter, this behavior is caused by the fact that annihilation operator "placed" in a part of the beam affects the whole beam profile. To underline the last statement, we present a one-dimensional beam profiles in Fig. 4e. The 1D-profile region is selected by the colored lines in Fig. 4a–d. It is easy to see, that a profile under photon subtraction repeats the initial profile, but it is twice higher.

In conclusion we report a demonstration of the "quantum vampire" effect in its initial representation: the photon annihilation in a part of the beam leads to the photon subtraction in the whole beam, so it doesn't change the beam profile (doesn't cast a shadow), but may change its brightness. Particularly, in the case of the thermal state as the input, the beam power is doubled. We should underline, that this effect is based on conditional implementation of the photon annihilation i. e. on the correlation measurements. Moreover, the input beam should be a single-mode. In some sense, the "quantum vampire" effect is opposite to the ghost imaging technique [28], where initial beam is multimode: an image, invisible in single counts, can be restored from coincidence measurement.

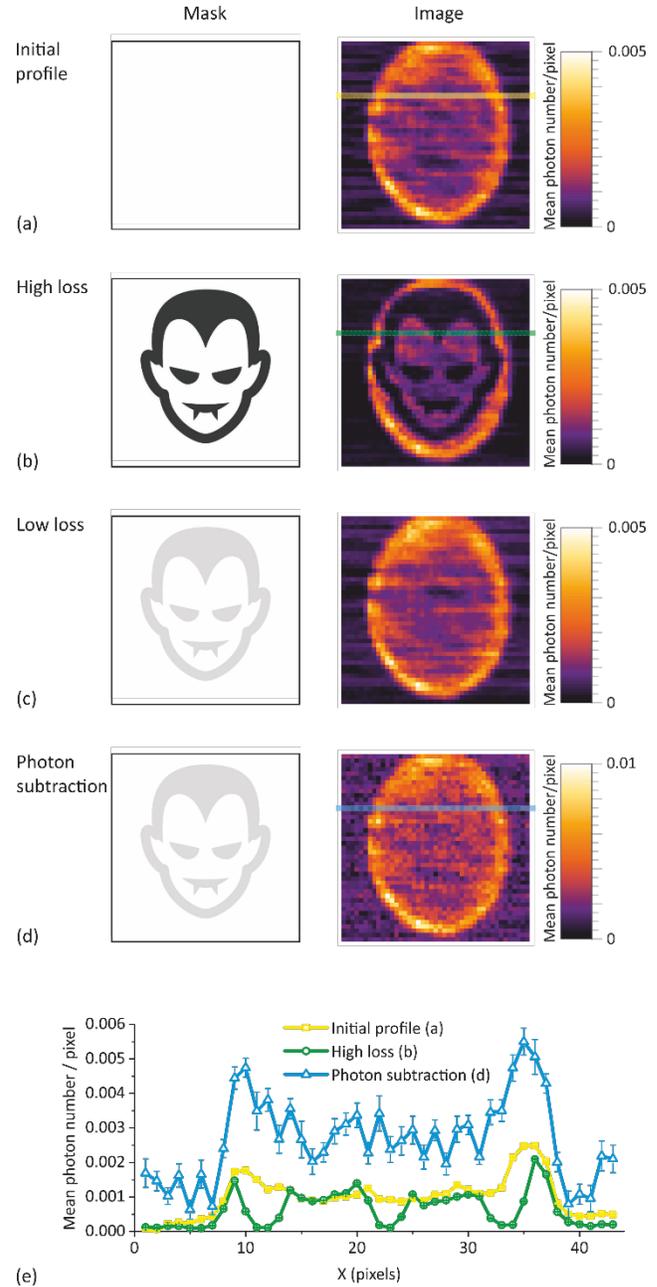

Fig. 3. The experimental results. (a) Initial beam profile; (b) high losses in the channel, (c) low losses in the channel, (d) conditional profile corresponded to photon subtraction. (e) One-dimensional profiles: yellow squares – initial beam, green circles – high losses in the channel, blue triangles – photon subtraction.

**Funding.** Russian Foundation of Basic Research (17-02-00790 A), Foundation for the Advancement of Theoretical Physics and Mathematics "BASIS" (Postdoc 17-13-334-1)

**Acknowledgment**. Authors express their gratitude for support to Vladimir Ermakov and Maria Rezantseva.